\documentclass[10pt, conference]{IEEEtran}

\usepackage{amssymb}
\usepackage{amsfonts}
\usepackage{amsmath,amsthm}
\usepackage[dvips]{graphicx}
\usepackage{cite}
\usepackage{balance}
\usepackage{color}
\usepackage{xcolor}
\usepackage{epstopdf}
\usepackage{algorithm}
\usepackage{algorithmic}
\usepackage{bm}
\usepackage{wrapfig}
\usepackage{subcaption}
\usepackage{url}
\usepackage{multirow}
\usepackage{mathtools}
\usepackage{tcolorbox}
\usepackage{enumitem}
\usepackage{array}
\usepackage{mdframed}

\def\BibTeX{{\rm B\kern-.05em{\sc i\kern-.025em b}\kern-.08em
		T\kern-.1667em\lower.7ex\hbox{E}\kern-.125emX}}

\graphicspath{{Figures/}}

\def\1{\mathds{1}}

\IEEEoverridecommandlockouts

\begin{document}
\title{Measurement-driven Analysis of an Edge-Assisted Object Recognition System}
\author{
	\IEEEauthorblockN{Apostolos Galanopoulos\IEEEauthorrefmark{1}, V\'ictor Valls\IEEEauthorrefmark{2}, George Iosifidis\IEEEauthorrefmark{1}, Douglas J. Leith\IEEEauthorrefmark{1}}
	
	\IEEEauthorblockA{\IEEEauthorrefmark{1} School of Computer Science and Statistics, Trinity College Dublin, Ireland}
	\IEEEauthorblockA{\IEEEauthorrefmark{2} Department of Electrical Engineering, and Institute for Network Science, Yale University, USA}
}
\maketitle

\begin{abstract}
We develop an edge-assisted object recognition system with the aim of studying the system-level trade-offs between end-to-end latency and object recognition accuracy. We focus on developing techniques that optimize the transmission delay of the system and demonstrate the effect of image encoding rate and neural network size on these two performance metrics. We explore optimal trade-offs between these metrics by measuring the performance of our real time object recognition application. Our measurements reveal hitherto unknown parameter effects and sharp trade-offs, hence paving the road for optimizing this key service. Finally, we formulate two optimization problems using our measurement-based models and following a Pareto analysis we find that careful tuning of the system operation yields at least 33\% better performance for real time conditions, over the standard transmission method.
\end{abstract}

\vspace{1mm}

\section{Introduction} \label{sec:intro}

Edge-assistance will most likely be a key component of future latency-critical and computationally-demanding mobile applications such as video analytics and Tactile Internet services~\cite{killer_app,Zhang18}. Augmented Reality~\cite{edge_assisted} and real time object recognition~\cite{Glimpse} are examples of such services that can benefit from the computational power of a nearby edge server, since mobile devices are too slow to timely perform the required computations. 
Nevertheless, the practical performance benefits of such edge architectures remain unexplored. On the one hand, data transmissions are added to the service delay. On the other hand, the quality and execution delay of analytics is affected by the volume of the transmitted data, as well as the complexity of the algorithm running on the edge server. 

\begin{figure}[t]
	\centering
	\includegraphics[width=\columnwidth]{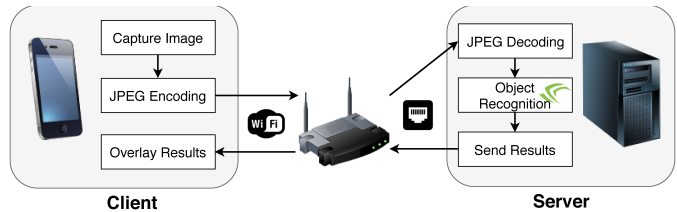}
	\caption{Schematic of edge-assisted object recognition system.}
	\label{fig:system}
	\vspace{-1mm}
\end{figure}

In this paper we investigate this issue experimentally, by building the edge computing system illustrated in Fig.~\ref{fig:system}. We develop a real-time object recognition system, as a representative of the plethora of emerging visual-aided services, e.g. video stream analytics, mobile augmented reality, etc. A mobile handset (client) captures camera images and transmits them to an edge server for processing; the server uses a deep neural network (NN) to detect and classify objects in the images; and sends the output to the handset which overlays this information on the screen.
We built the above system using an Android application and a state-of-the-art deep learning network running on GPU hardware for the server. We use a high performance 802.11ac wireless link for communication between the handset and the server, which features technology likely to persist in future small cells\footnote{We use MU-MIMO/OFDM and channel aggregation at the PHY layer, and employ packet aggregation at the MAC layer to reduce framing/signaling overheads.}, hence making our results relevant to a range of systems.

Our goal is to understand the system-level trade-offs between end-to-end (E2E) latency and object recognition accuracy, and propose specific solutions that can improve the performance of the system.
We firstly show that the degree of image compression and deep learning NN input size are  key parameters affecting both performance metrics. In particular, the use of more aggressive image compression saves on communication latency between client and server (since the transmitted image file is smaller), but at the cost of reduced object recognition accuracy. While the impact of image degradation due to noise or blur on recognition accuracy has started to receive attention in the deep learning literature~\cite{dodge2016}, the impact of compression on accuracy remains relatively poorly understood. Furthermore, a large NN size will improve recognition performance at the cost of higher execution delay at the server, hence increasing E2E latency. To the best of our knowledge, the trade-off between E2E latency and recognition accuracy for the above parameters, has not previously been explored.

We focus our effort in designing wireless transmission interventions that further improve the communication delay of the system. Such interventions have not yet received significant attention by the edge computing literature, as most efforts have been devoted to minimizing computation delays\cite{DeepMon,DeepDecision,AAY18}. This delay source however, is of critical importance to low latency services, and hinders their ability to achieve real time performance, e.g.~\cite{Overlay,Glimpse}. 
We show that \emph{transmit time can be reduced by up to 65\%} by sending the images as short back-to-back bursts of UDP packets. We also find that the client Network Interface Controller (NIC) powersave can incur substantial transmit latency and, hence, \emph{smarter sleep mode adaptation can further decrease latency by up to 60\%}.

Finally, we model the different sources of delay in our system, and the obtained accuracy, as functions of the NN size and encoding rate using our measurements. We illustrate the use of the developed model to highlight optimal trade-offs between E2E latency and system object detection accuracy. Moreover, we show that the use of smart wireless transmission techniques employed, can nearly double the system performance along the Pareto-optimal curve of accuracy vs frame rate.
Our main contributions are as follow. 
\begin{itemize}[leftmargin=*]

\item	We build the edge architecture of Fig.~\ref{fig:system}, where the image encoding rate and input NN size are tunable parameters. 
	
\item	We tailor the system design, with wireless transmission interventions (Transport layer, MAC aggregation, device wake-up), reducing the communication delay to just 2-6 ms.

\item   Using the system, we explore the impact of image encoding quality and NN size on the delay and recognition accuracy. Extensive experiments reveal sharp trade-offs between these two performance criteria.

\item  We collect a wealth of measurements and use them to build statistical models for the performance metrics of interest. These can be used in order to tailor the system operation based on the needs of the client, e.g. maximize accuracy for a minimum perceived frame rate.

\end{itemize}

\textbf{Paper Organization}. In Sec.~\ref{sec:system} we describe the system architecture and the evaluation scenario. In Sec.~\ref{sec:latency_accuracy} we measure the impact of the image encoding and NN size on the E2E latency, and present our design choices for reducing the transmission delay. In Sec.~\ref{sec:trade-offs} we analyze the inherent latency-accuracy trade-off, while in Sec.~\ref{sec:optimiz} we use our measurements to obtain analytical models for delay and accuracy. Finally, Sec.~\ref{sec:related} presents a discussion of the related work, while Sec.~\ref{sec:conclusion} concludes the paper.


%
\section{Preliminaries} \label{sec:system}

\subsection{Hardware \& Software Setup}


We developed an Android application that captures images through the handset's camera, carries out JPEG encoding and then transmits the compressed images to an edge server for processing. The server software (written in C/C++) decompresses and pre-processes the images, and submits them to the deep learning neural network (NN) which is implemented using a GPU-optimized framework. The results, i.e., the bounding boxes and labels, are then sent back to the client handset and overlaid on the displayed image. 

Object recognition is performed by YOLO~\cite{yolov3}, a state-of-the-art deep learning detector implemented on darknet, an open source framework that supports GPU computations via cuda. It takes an $n\!\times\!n$ array of image pixels as input, with each pixel being a float value, and down-samples by $32$ to give an $n/32$ grid. Then, each grid cell proposes bounding boxes and labels for any contained objects. These results are filtered to generate the output consisting of a set of bounding boxes of recognized objects with their labels and respective confidence. 

We use different mobile devices to measure the effect of the end user's hardware on the system's performance: \textit{(i)} a Google Pixel 2 (default device), \textit{(ii)} a Samsung Galaxy S8, and \textit{(iii)} a Huawei P10 Lite. All devices are equipped with 802.11ac chipsets, and we will be using the Google phone unless stated otherwise. The edge server is connected via Ethernet to a WiFi router that serves as an access point (802.11ac, 5GHz) for the handsets\footnote{The edge server is a 3.7 GHz Core i7 PC equipped with 32GB of RAM and a GeForce RTX 2080Ti GPU; and the router is the ASUS RT-AC86U.}, see Fig.~\ref{fig:system}.


\subsection{The Need for Edge Server Offload}
We investigated first the viability of running YOLO on the handset by cross-compiling darknet, but found that the running times were excessive (on the order of minutes). Use of a cut-down version of YOLO, referred to as TinyYOLO~\cite{yolov3}, was also investigated. The running time was around 1s per image, substantially faster than with the full YOLO network but still very slow compared to the server. Note also that the speedup of TinyYOLO is obtained at the cost of significantly reduced object recognition accuracy, and supports only a small subset of object types. Our tests convey the same message as previous studies \cite{DL_offloading,LAVEA}, namely confirm the \emph{necessity for offloading the object recognition task to a powerful server, if low latency operation is to be obtained}.

\subsection{Evaluation Scenario}

To evaluate the system performance we used the extensive COCO dataset~\cite{COCO} which covers a wide range of images and objects, and includes ground truth for each image (object locations and labels within each image). For quantifying performance, we used the Average Precision (AP) and Average Recall (AR) metrics\footnote{AP is the ratio $T_p/(T_p+F_p)$ while AR is the ratio $T_p/(T_p+F_n)$, with $T_p$ being the true positive detections, $F_p$ the false positive and $F_n$ the false negative detections. The results are averaged over all objects classes.} for a range of Intersection-over-Union (IoU) values. Detection is considered successful when the ratio of the overlapping area between the detected object and the ground truth, over their respective union area, is higher than an IoU value of 0.5. COCO further breaks precision and recall metrics down by whether objects are large, medium or small. YOLO is known to perform poorly on small objects and so we focus on large and medium objects.

To use the COCO images we connected the phone to a server via a USB cable and a Python script on the server sends commands to the phone using the Android Debug Bridge (adb). The server initiates the client application through adb and configures the system parameters for the experiment (e.g., the JPEG compression level). Then it iterates over 5000 images from the COCO validation set, sending them one-by-one to the phone through cable. The phone transmits each image to the server through the wireless interface, as if they were images captured by its camera, receives the server response over WiFi and passes this back over the USB cable for logging.

\section{System End-to-End Latency} \label{sec:latency_accuracy}


Our first goal is to measure each of the different delay components involved in the procedure, and investigate how they are affected by the encoding rate $q$ and NN size $n$, but also by the network set up (from the transport, to data link and physical layer). Based on our findings we propose and evaluate network design choices that speedup the task completion.

\subsection{Encoding Delay ($\,T_{enc}\,$)}

The handset application converts its camera images to JPEG format before transmission to the server. We use JPEG as it is widely adopted and supported by the Android API. While image encoding is a typical step in such systems, its impact on the performance of edge-assisted object recognition has not received attention, with only few exceptions~\cite{Zhang18}. JPEG is a lossy format and its compression is decided by the \emph{encoding rate} $q$. Note that we rely on the terminology of the compression library we employed in our system\footnote{For jpeg compression (through quantization) we used the Android library: \url{https://developer.android.com/reference/android/graphics/YuvImage}.} and define $q\!\in\![10,100]$ as the percentage ratio of compressed image size over its actual size, where $q\!=\!100$ for an uncompressed image.

At higher encoding rates, the number of discrete cosine transform coefficients that represent the JPEG image is larger, leading to an expected increase in the encoding delay. Indeed, Fig.~\ref{fig:encoding} (upper plot) shows the encoding delay $T_{enc}$ vs. the encoding rate $q$. It can be seen that $T_{enc}$ grows from 5ms to 11ms as $q$ increases from 25\% to 100\%. This has also impact on the size of the compressed image, see Fig.~\ref{fig:encoding} (lower plot).

\subsection{Decoding and Pre-processing Delay ($\,T_{dec}\,$)}

\begin{figure}[t]
	\centering
	\begin{subfigure}{.48\linewidth}
		\includegraphics[scale=0.2]{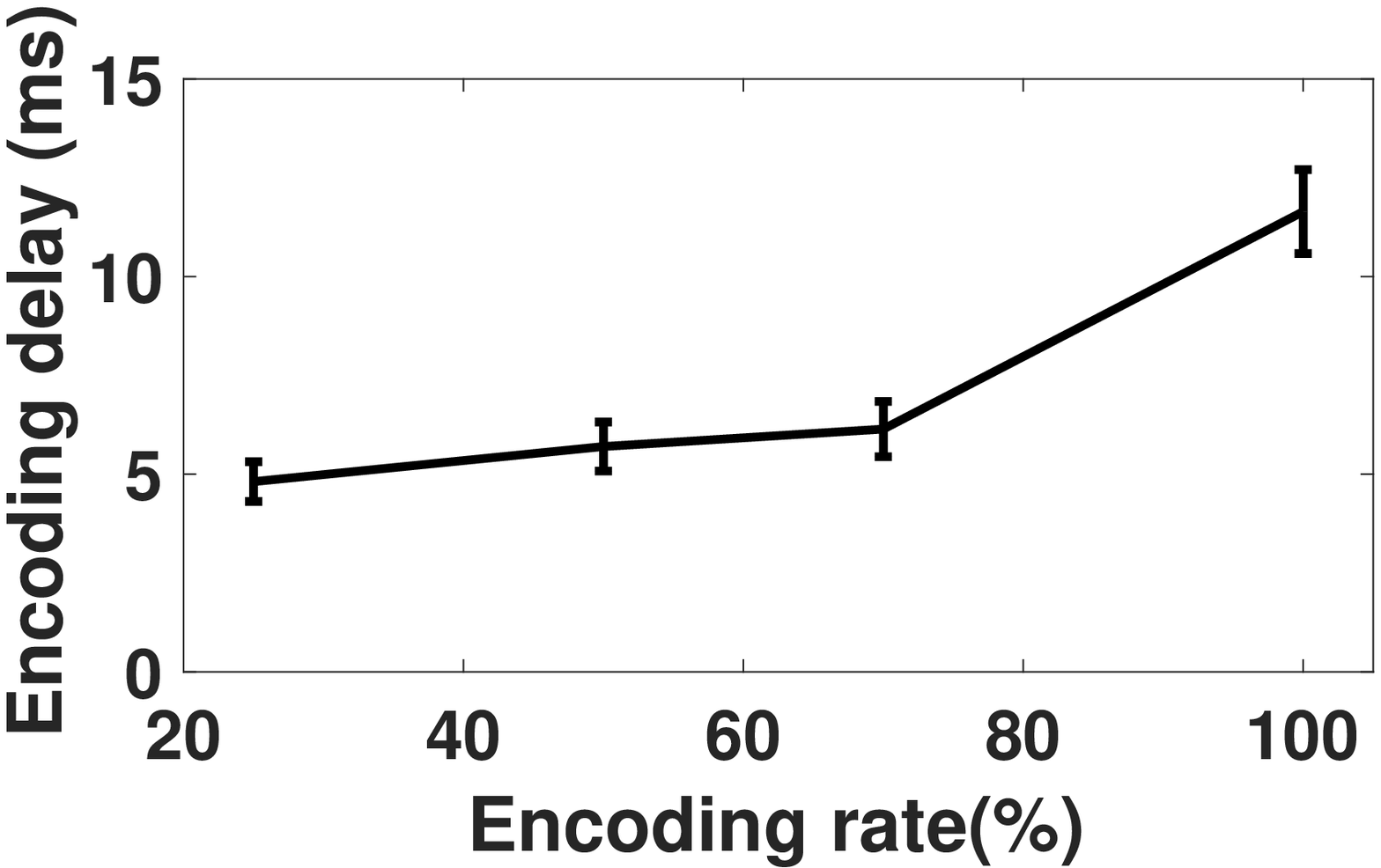}\\
		\includegraphics[scale=0.2]{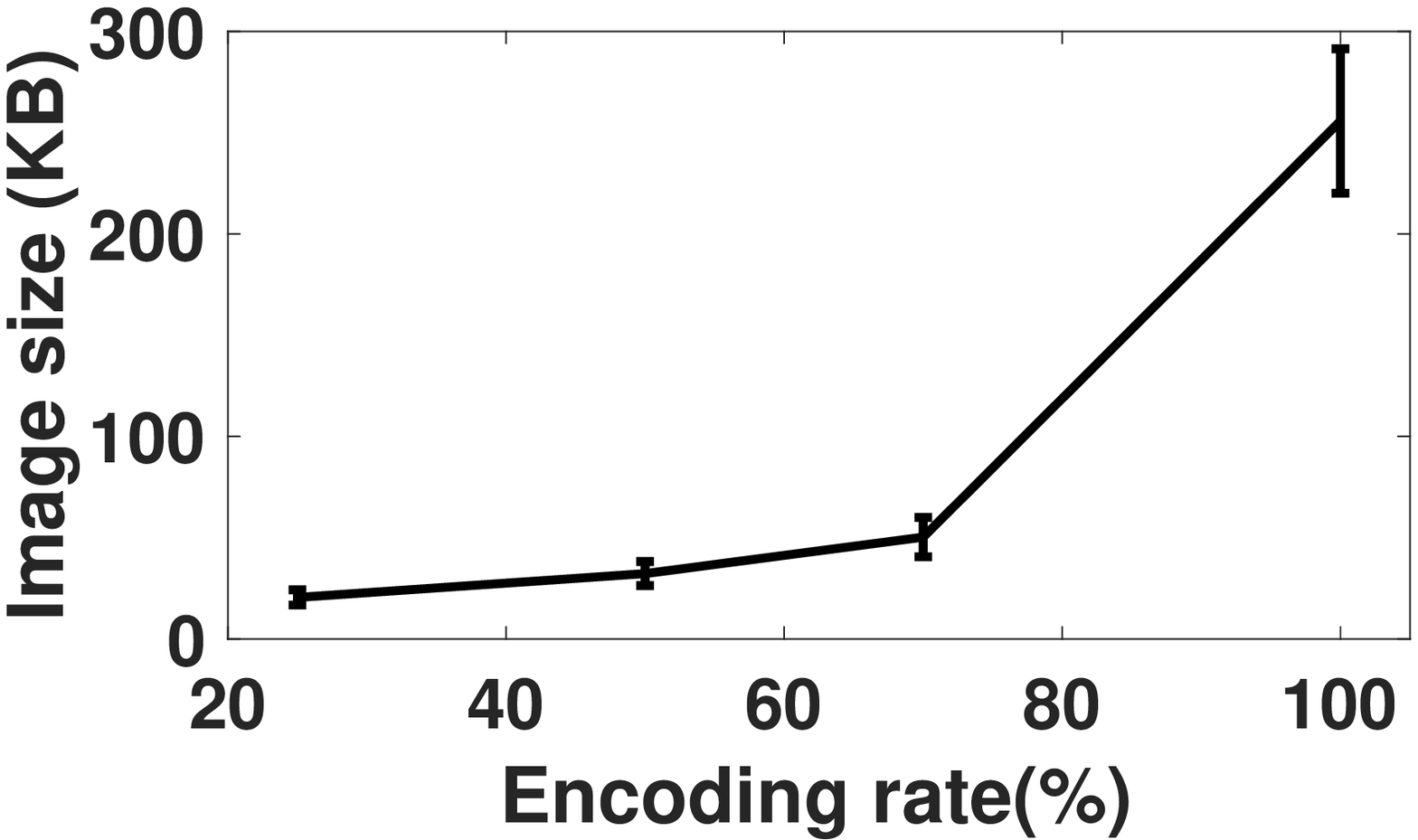}
		\caption{}
		\label{fig:encoding}
	\end{subfigure}
	\begin{subfigure}{.48\linewidth}
		\includegraphics[scale=0.26, trim=2cm 0cm 2cm 2cm]{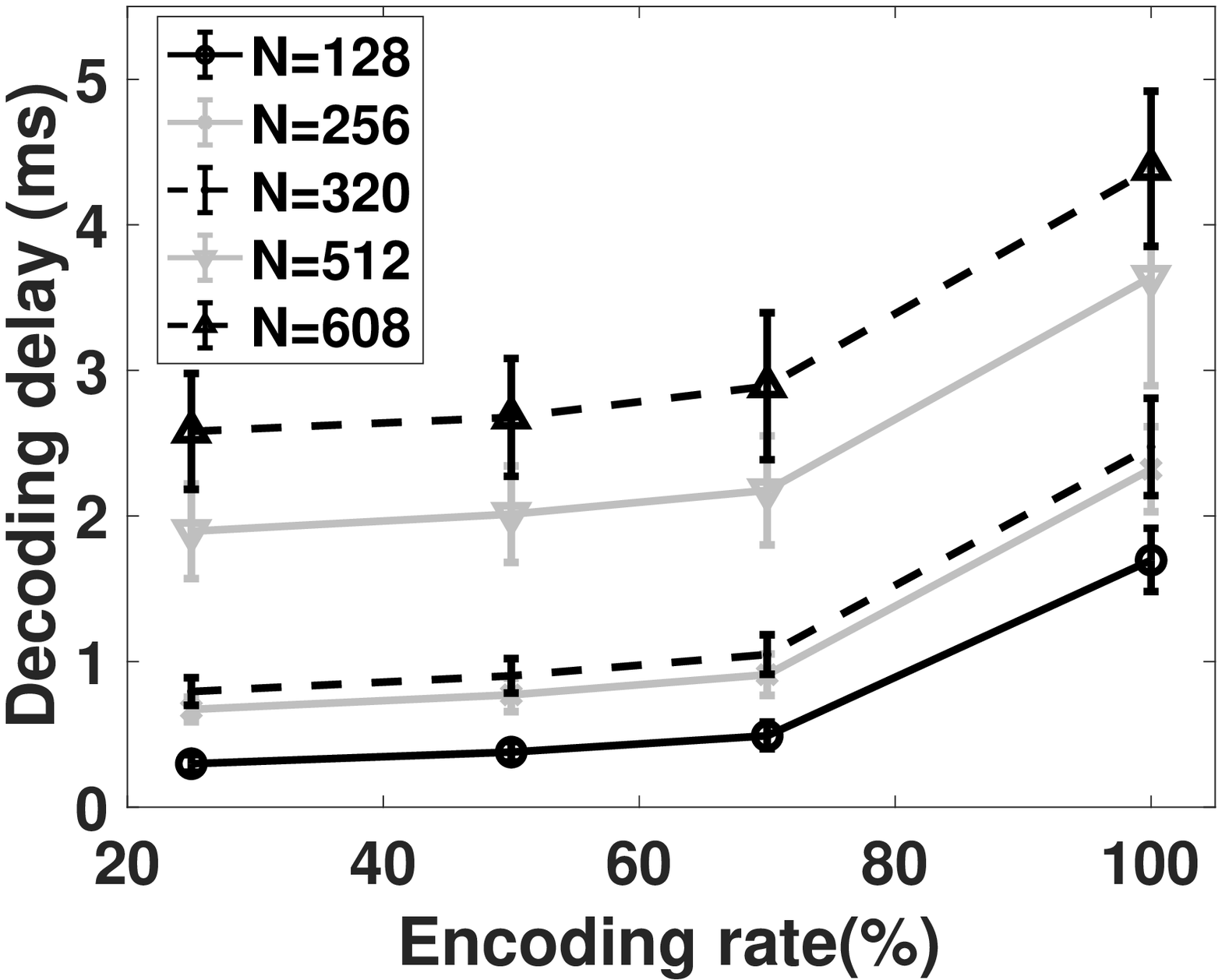}
		\caption{}
		\label{fig:decoding2}
	\end{subfigure}
	\vspace{-1mm}
	\caption{Time used for: (a) JPEG encoding, (b) decoding and preprocessing, vs encoding rate $q$. Results are averaged for the entire COCO library (5000 images).}
\end{figure}

Upon receiving an image, the server \emph{(i)} decompresses it to obtain an RGB image; \emph{(ii)} re-samples/pads the image to match the input size $n$ of the deep learning network; \emph{(iii)} rotates the image to compensate for the handset camera orientation; and \emph{(iv)} converts the pixel values from 0-255 integers to 0-1.0 floats. Our profiling indicates that \emph{most of this processing is limited by memory resources rather than CPU}. Hence, in our implementation we execute steps \emph{(i)} and \emph{(ii)} jointly so as to minimize memory movements and maximize scope for in-processor caching. And similarly we designed our implementation to execute simultaneously steps \emph{(iii)} and \emph{(iv)}. Contrary to encoding delay, this part of the processing depends both on the encoding rate and the NN size. Fig.~\ref{fig:decoding2} plots measurements of the processing time vs. $q$ and $n$. Observe that when $q\!\leq\!75$ the latency is largely insensitive to $q$, i.e., it is dominated by the preprocessing steps other than image decompression. Similarly, the NN size $n$ affects significantly $T_{dec}$ only when it is very large (notice the sudden increase when $n\!\geq\!512$). As we will see later, these findings create opportunities for optimizing the overall system operation.

\subsection{Transmission Delay ($\,T_{tx}\,$)}
Next, we investigate the network impact on the task delay, and propose specific solutions that can effectively halve this time. First, note that the size of the transmitted images vary between 20--250KB, corresponding to roughly 13--166 packets (each 1500B long). In contrast, the server response contains object bounding boxes and typically fits into a single packet. Hence, the network transmission delay is dominated by the time taken to transmit the image and we expect that this will increase with the encoding rate $q$.

\begin{figure}
	\centering
	\begin{subfigure}{.48\linewidth}
		\includegraphics[scale=0.22]{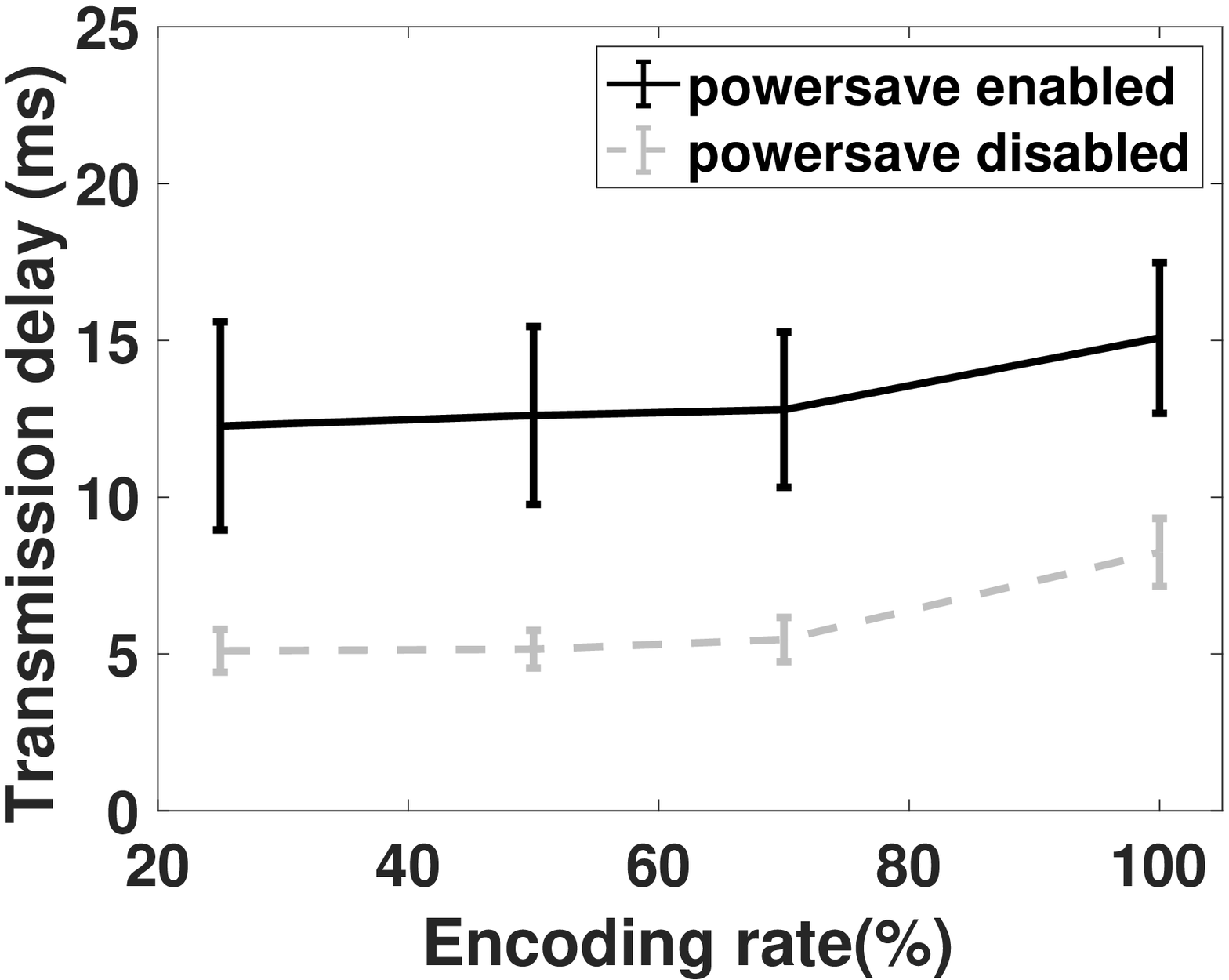}
		\caption{}
		\label{fig:tcpdelay}
	\end{subfigure}
	\hspace{0.1cm}
	\begin{subfigure}{.48\linewidth}
		\includegraphics[scale=0.215]{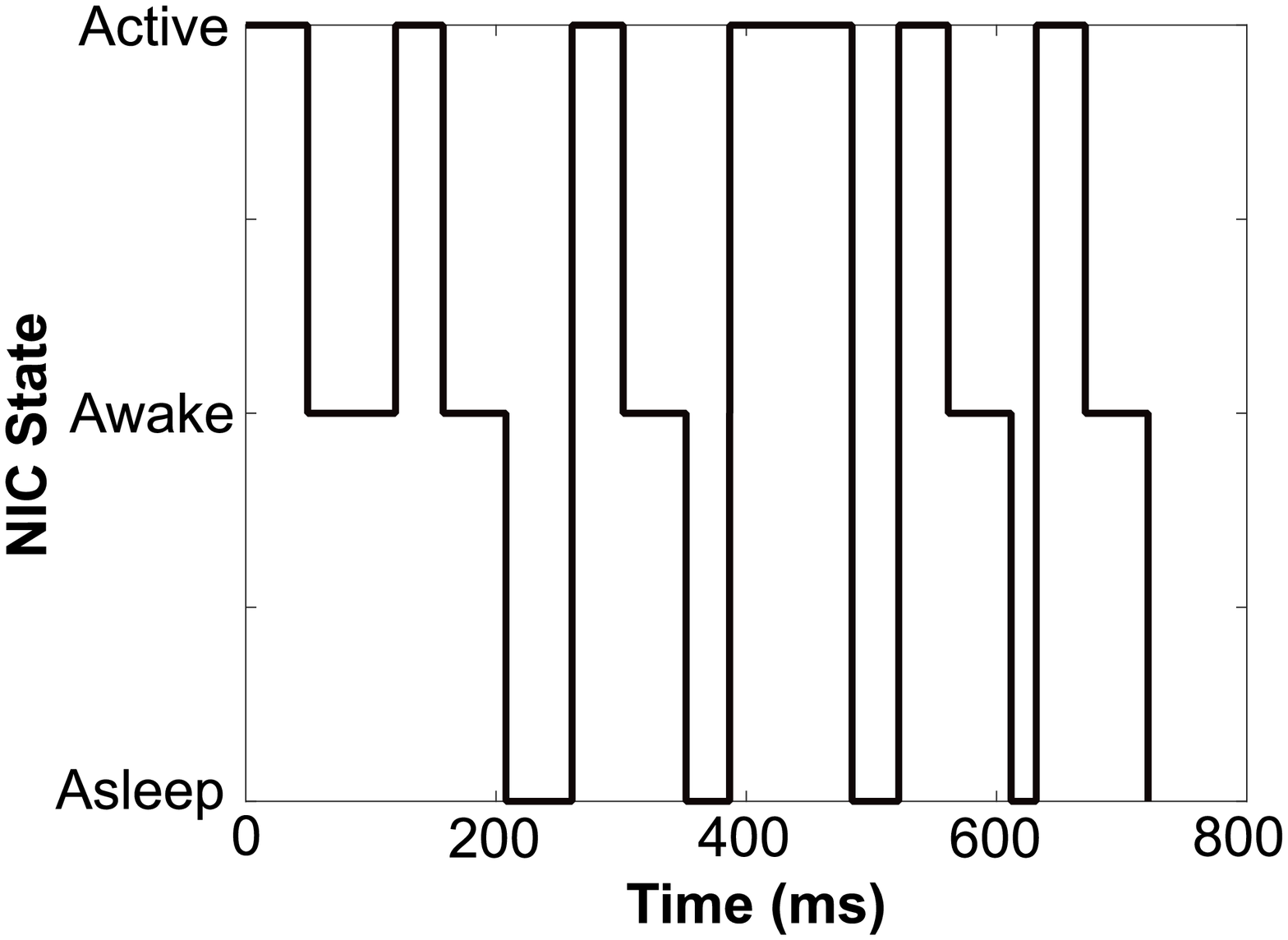}
		\caption{}
		\label{fig:powersave}
	\end{subfigure}
	\vspace{-1mm}
	\caption{(a) Wireless transmission delay using TCP vs JPEG encoding rate, (b) example time history of the NIC state on the mobile handset when power saving is enabled. }
\end{figure}

The solid line in Fig.~\ref{fig:tcpdelay} plots the transmission delay vs. $q$. This delay includes the time needed to send the image to the server and the time for transmitting back the response. The measurements are when TCP is used with default Android and Linux settings, i.e., Cubic congestion control and dynamic socket buffer sizing. As expected, the delay tends to increase with the JPEG quality (for larger $q$). However, when $q\!<\!80$ the delay is relatively insensitive to the encoding rate. \emph{Further investigation reveals that this insensitivity is mainly caused by two factors}. Firstly, the handset's power management aggressively puts the NIC into sleep mode, and this induces a delay to wake the NIC when transmission or reception restarts. Secondly, the dynamics of TCP congestion control mean that it takes multiple round-trip times to transmit all image packets. Next, we propose solutions for these two issues.

\subsubsection{Handset NIC Wake-from-Sleep Latency}
When entering sleep mode, the handset's 802.11 NIC sends a special flagging frame to the AP which buffers any packets awaiting transmission until the handset signals it has woken up. Fig.~\ref{fig:powersave} plots an example time history of the handset's NIC state derived by extracting these state transitions from tcpdump data\footnote{In our experiment a delay is inserted between input of each image to the android app to make the power-save behavior easier to see.}. Also indicated on Fig.~\ref{fig:powersave} are ``active'' periods where the NIC is awake and exchanges data with the server. Note that the NIC regularly enters a sleep state, waking up when the handset starts to send an image. As indicated by our measurements above, the handset can roughly predict when the next image transmission will occur. Namely, a new captured image is transmitted approximately after 5-10ms (time for its encoding), and this could be used to preemptively wake up the NIC.

\underline{Solution:} In order to investigate the potential latency gains of smart wake-up strategies, we adopted the cruder approach of using iperf to generate 1Mb/s of background UDP traffic from the server to the client, to keep the handset's wireless interface awake. The dashed line in Fig.~\ref{fig:tcpdelay} shows that the overall transmit delay is now decreased for all values of $q$, consistent with the handset NIC no longer having to be woken up for transmitting the image. \emph{The delay reduction is approximately 5ms for all encoding rates which corresponds to a reduction of 50\% in the wireless transmission delay.}

\subsubsection{Latency Caused By TCP Dynamics}
\begin{figure}
	\centering
	\begin{subfigure}{.48\linewidth}
		\includegraphics[scale=0.21]{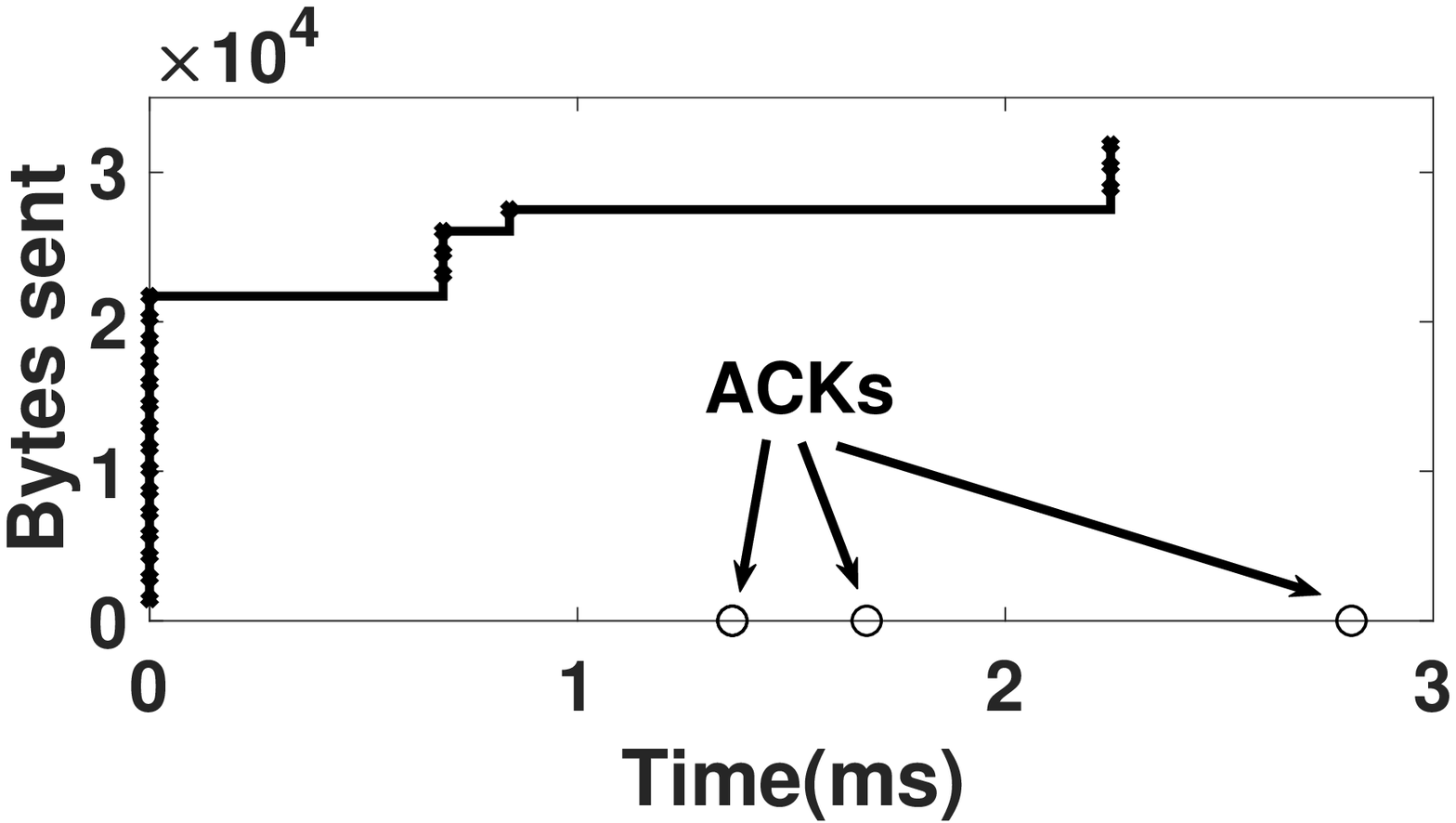}\\
		\includegraphics[scale=0.21]{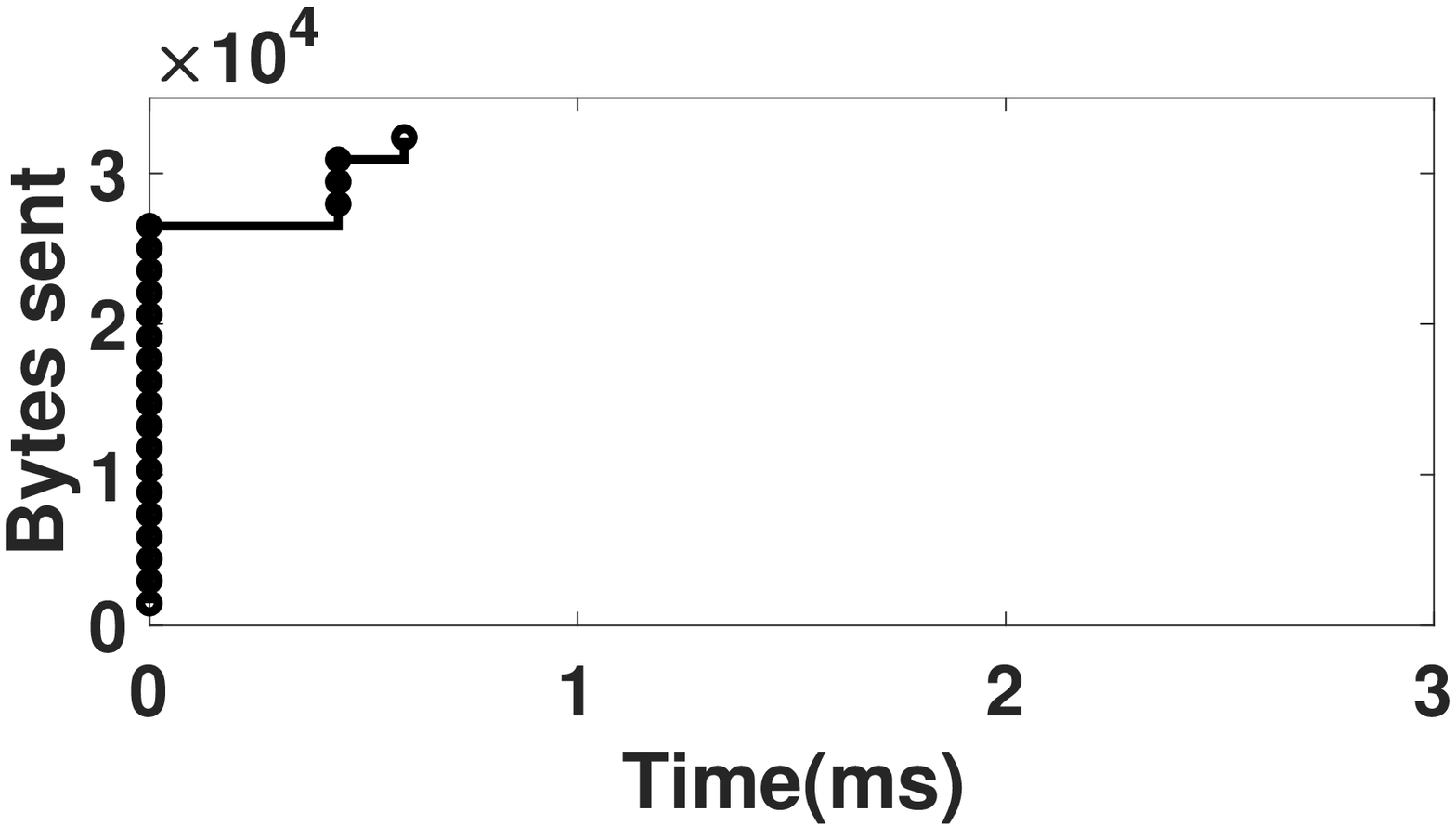}
		\caption{ }
		\label{fig:tcpvstime}\label{fig:udpvstime}
	\end{subfigure}
	\hspace{0.1cm}
	\begin{subfigure}{.48\linewidth}
		\includegraphics[scale=0.21]{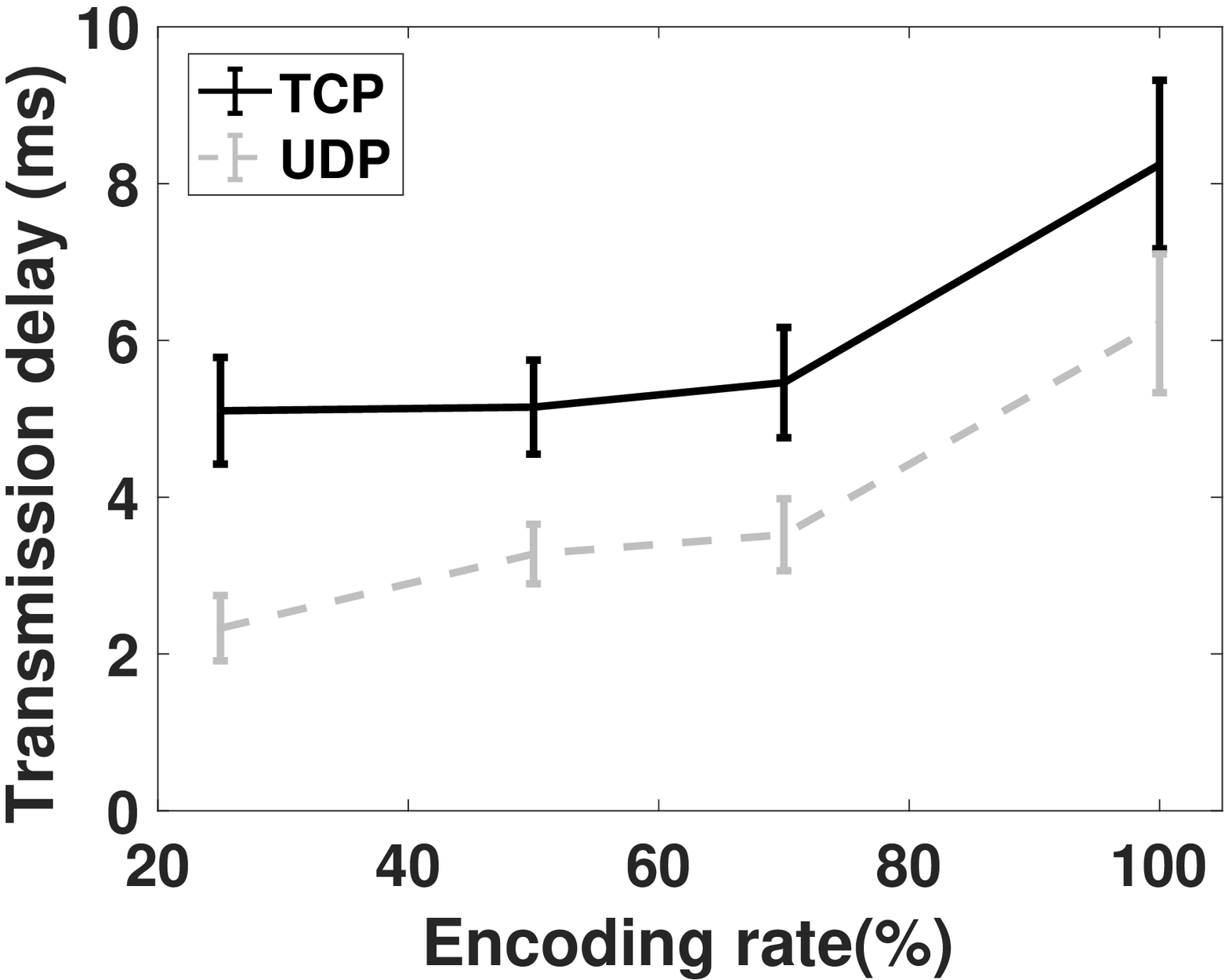}
		\caption{}
		\label{fig:tcp_udp_delay}
	\end{subfigure}
	\vspace{-1mm}
	\caption{(a) Time histories showing transfer of a compressed image from client to server using TCP (upper plot) and UDP (lower plot), markers indicate packet boundaries. (b) Wireless transmission delay for TCP and UDP vs JPEG encoding rate $q$ with mobile NIC power-save disabled.}
\end{figure}

The upper plot in Fig.~\ref{fig:tcpvstime} shows the time history when transferring an image using TCP. The connection is kept open and used for sending multiple images so that the overhead of the TCP handshake (SYN-SYNACK-ACK) is only incurred once (takes 4ms; not shown). The compressed image in this example is 31335B in size, and when the HTTP request header is added, it occupies 22 TCP packets\footnote{The payload of a 1500B TCP packet is 1448B including header overheads.}. Its transmission lasts 2.5ms and uses 4 MAC frames for data and 3 for TCP ACKs. On average, 5.5 TCP data packets are therefore sent in each MAC frame. Observe that the client needs to receive TCP ACKs before it can send the full image since the TCP congestion window (cwnd) limits the packets in flight to around 10 when starting a new transfer. Also, observe that there is contention between uplink and downlink due to the ACKs transmitted by the server.

\underline{Solution:} We explore the gains from removing uplink/downlink contention and the impact of TCP cwnd, by modifying the Android client and server to use UDP.  At the client side, an image is segmented and placed into a sequence of UDP packets which are then sent to the socket back-to-back to facilitate aggregation by the NIC. The lower plot in Fig.~\ref{fig:udpvstime} shows UDP measurements\footnote{Including the time needed to segment the image into UDP packets, so the values are comparable with the TCP data.} for transmission of the same image. Despite that UDP packets are fit within a single MAC frame (our system can aggregate up to 128 packets in 1 frame), we see that the transfer used actually 3 frames. Presumably this is due to the scheduling delays between the kernel and NIC, and the relative timing of channel access opportunities and packet arrivals. Nevertheless, we find that the data transfer time is now 0.8ms, i.e., 3 times faster than with TCP. Finally, Fig.~\ref{fig:tcp_udp_delay} plots measurements of the overall wireless transmission time (sending the image and receiving its response) for the full COCO data set when using TCP and UDP; and with mobile NIC power-save disabled. We find that using UDP packet bursting roughly halves the transmit time for all JPEG encoding rates. 

Concluding, in this subsection we showed that \emph{tailored transmission strategies, such as smart NIC power-saving and using UDP with packet bursting, reduce the transmit time to around 5ms}. This improvement is hugely important given the targeted E2E latency budgets.\footnote{To achieve real time frame update rates, such as 30fps, the available total latency budget is only 33ms.}

\begin{figure}[t]
	\centering
	\begin{subfigure}{.78\linewidth}
		\centering		
		\includegraphics[scale=0.23]{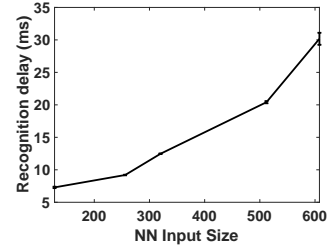}
	\end{subfigure}
	\caption{Server recognition delay ($T_{dl}$), for different NN size.} 		\label{fig:yolo_delay}
	\vspace{-1mm}
\end{figure}
 
\begin{figure}[t]
	\centering
	\includegraphics[scale=0.55]{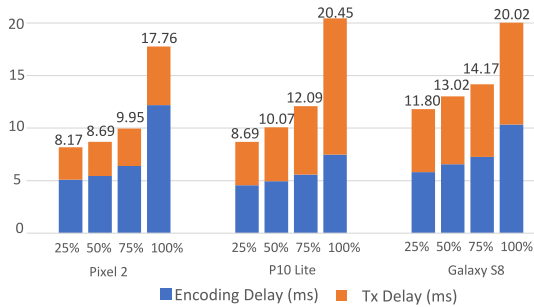}	
	\caption{Edge device delay comparison.}
	\vspace{-2mm}
	\label{fig:devices}
\end{figure}

\subsection{Recognition Delay ($\,T_{dl}\,$) and Impact of Handheld}

YOLO outputs the coordinates of the image's detected objects along with their labels. The recognition delay $T_{dl}$ depends on the NN size, and our measurements in Fig.~\ref{fig:yolo_delay} show that it increases, roughly, quadratically with $n$. Other works have reported similar findings, e.g., see \cite{DeepDecision, DL_offloading}, but the delays are quite higher than our results, presumably due to the usage of older GPU hardware. Furthermore, DeepMon~\cite{DeepMon} proposes NN optimizations on the mobile devices that reduce the delay at about 1sec for YOLO, but it is still worse than our system's performance. These values may vary from system to system, but we expect qualitatively the trend to persist.

Similarly, we suspect that the handset hardware affects only slightly (i.e., quantitatively) the results. To verify this, we repeat our experiments with 2 additional mobile devices. The delays that are directly related to the handset device, and may vary due to the different hardware specifications, are the encoding and transmission delay. Fig.~\ref{fig:devices} plots the total encoding and transmission delay measured for the 3 devices (Pixel 2, P10 Lite, Galaxy S8) for each encoding rate $q$ (averaging all dataset images). We find that compared to the Pixel 2, the other 2 devices are slightly faster in image encoding, but also slower in transmitting. Such differences might likely arise due to the different chipsets/firmware implementations. Observe however, that the roughly quadratic increase of both delay components persists across all devices as $q$ increases. Hence, qualitatively the results hold for different hardware.

\begin{figure*}[t]
	\centering
	\begin{subfigure}{.19\linewidth}
	\includegraphics[scale=0.28,trim={0 0 2cm 0},clip]{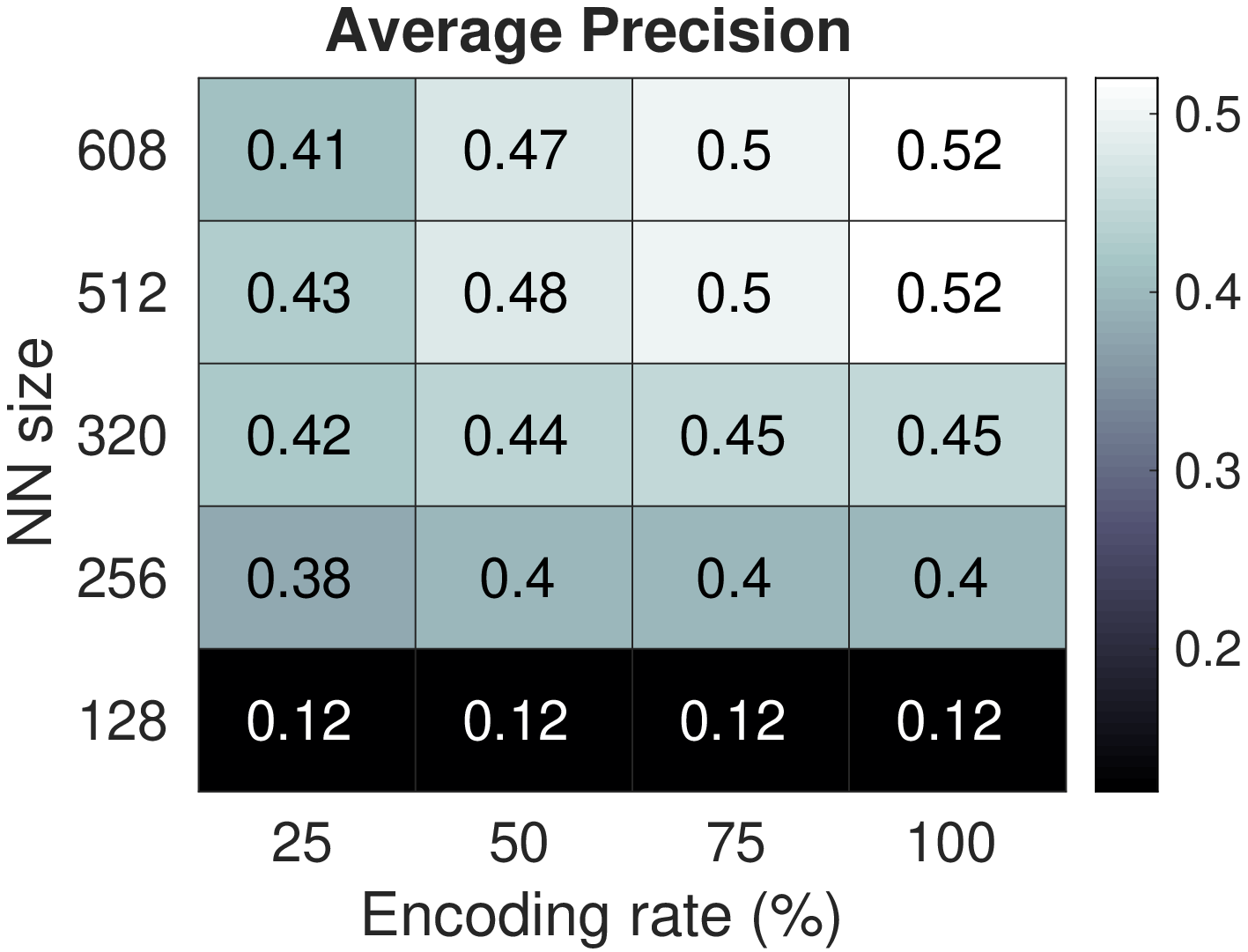}
	\caption{}
	\label{fig:precision}
	\end{subfigure}
	\begin{subfigure}{.19\linewidth}
	\includegraphics[scale=0.28]{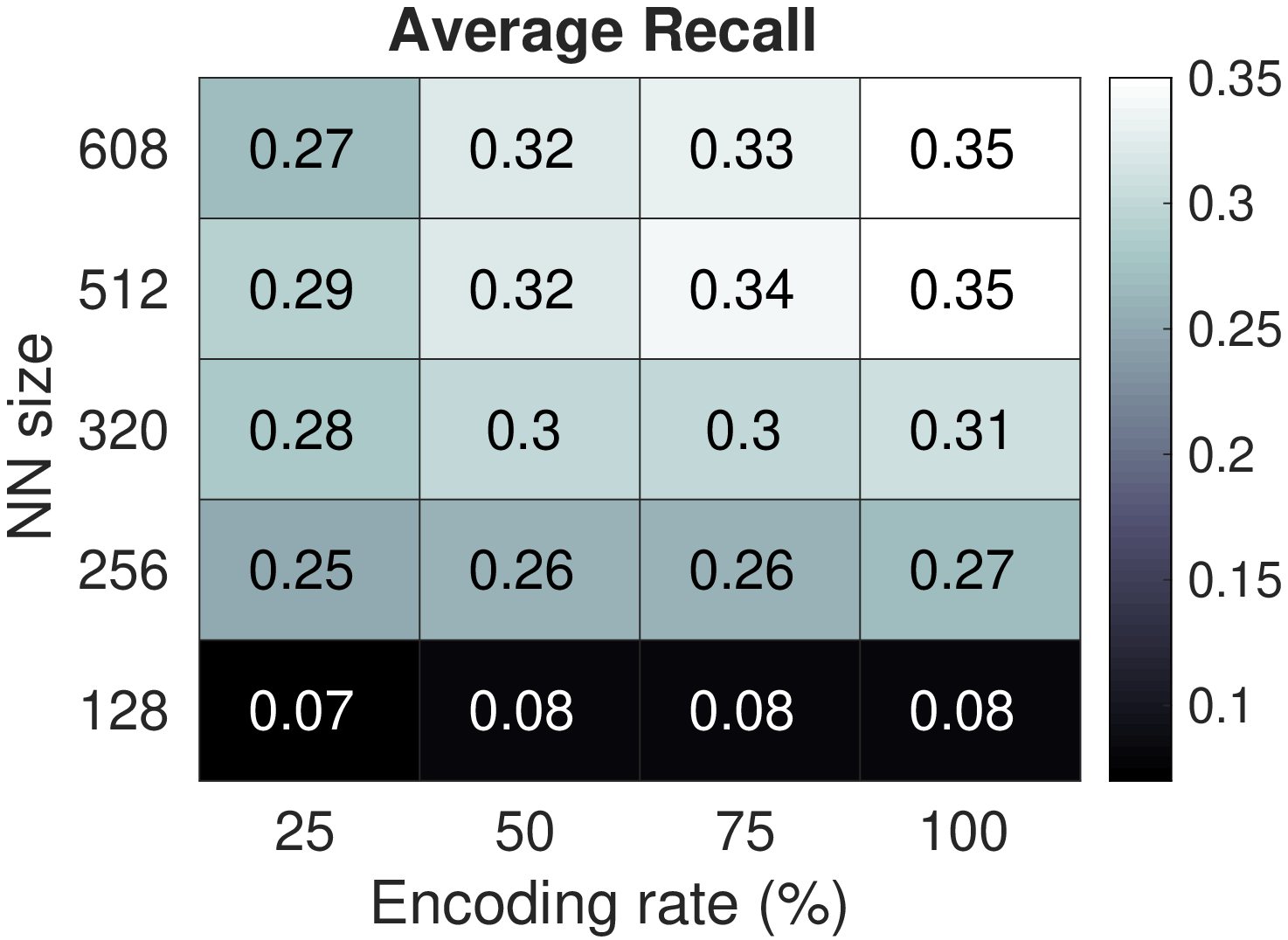}
	\caption{}
	\label{fig:recall}
	\end{subfigure}
	\begin{subfigure}{.19\linewidth}
	\includegraphics[scale=0.26,trim={.2cm 0 0 0},clip]{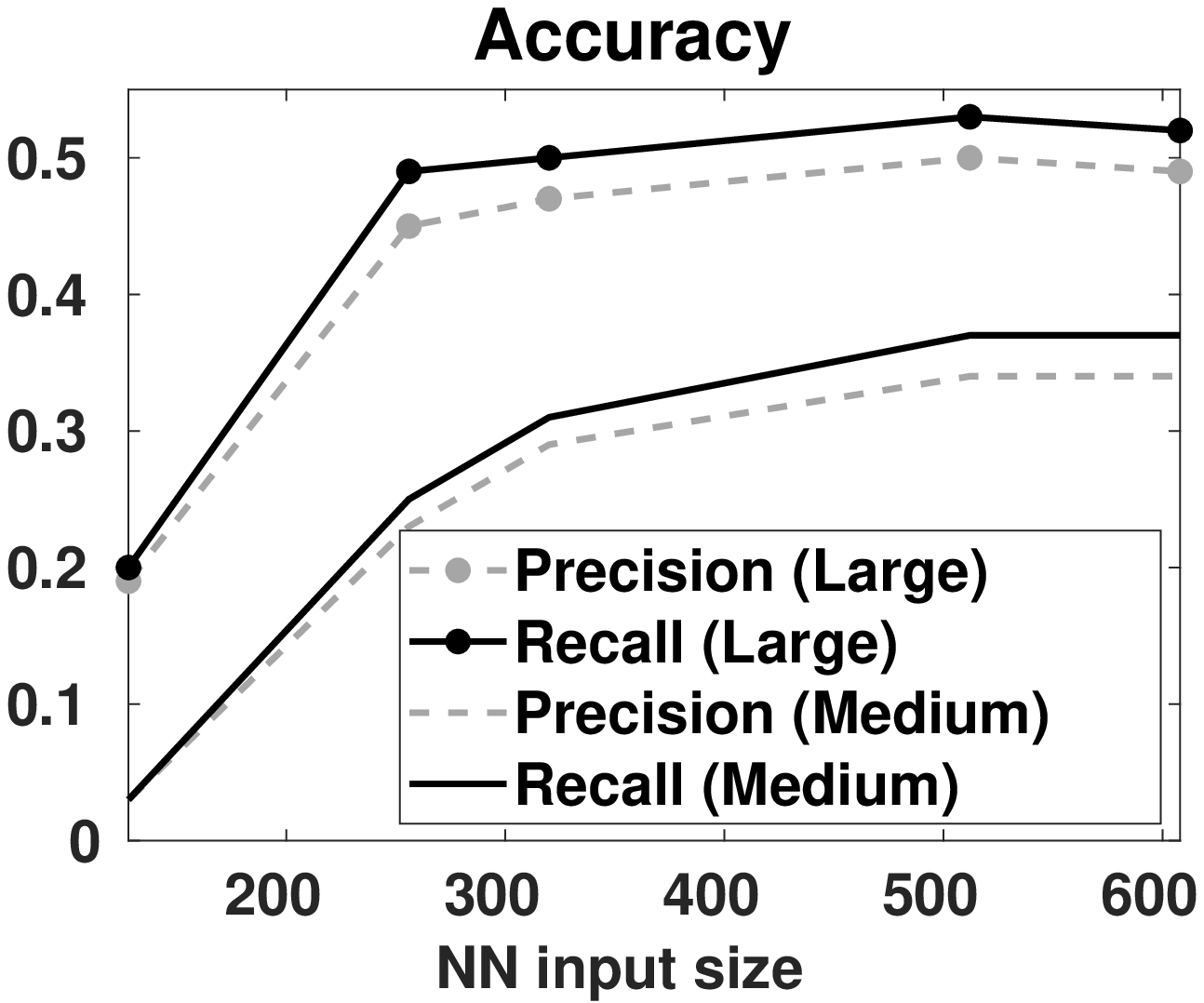}
	\caption{}
	\label{fig:accuracy_vs_N}
	\end{subfigure}
\,
	\begin{subfigure}{.19\linewidth}
	\includegraphics[scale=0.26,trim={.2cm 0 0 0},clip]{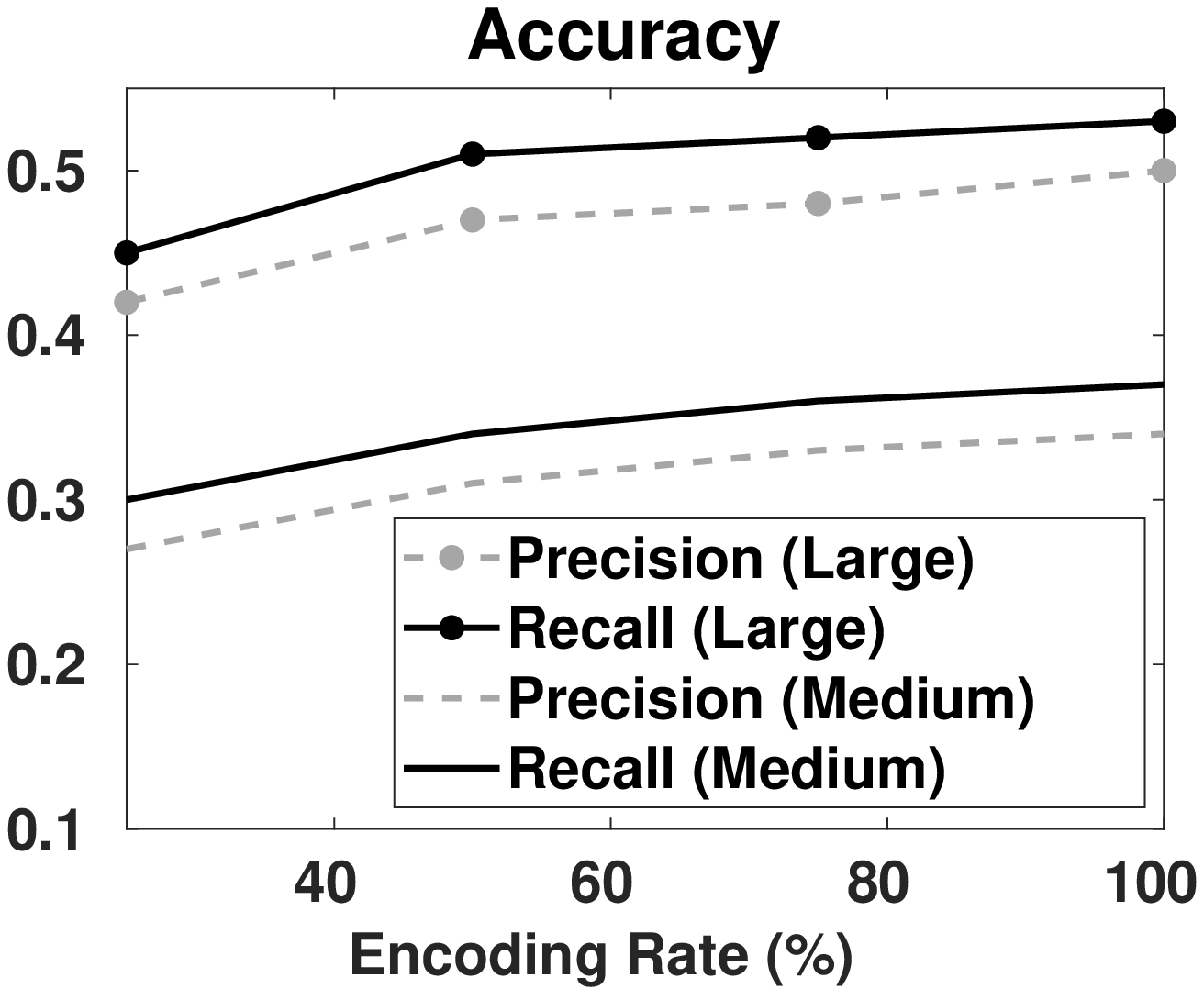}
	\caption{}
	\label{fig:accuracy_vs_Q}
	\end{subfigure}
	\begin{subfigure}{.2\linewidth}
	\includegraphics[scale=0.29,trim={.6cm 0 0 0},clip]{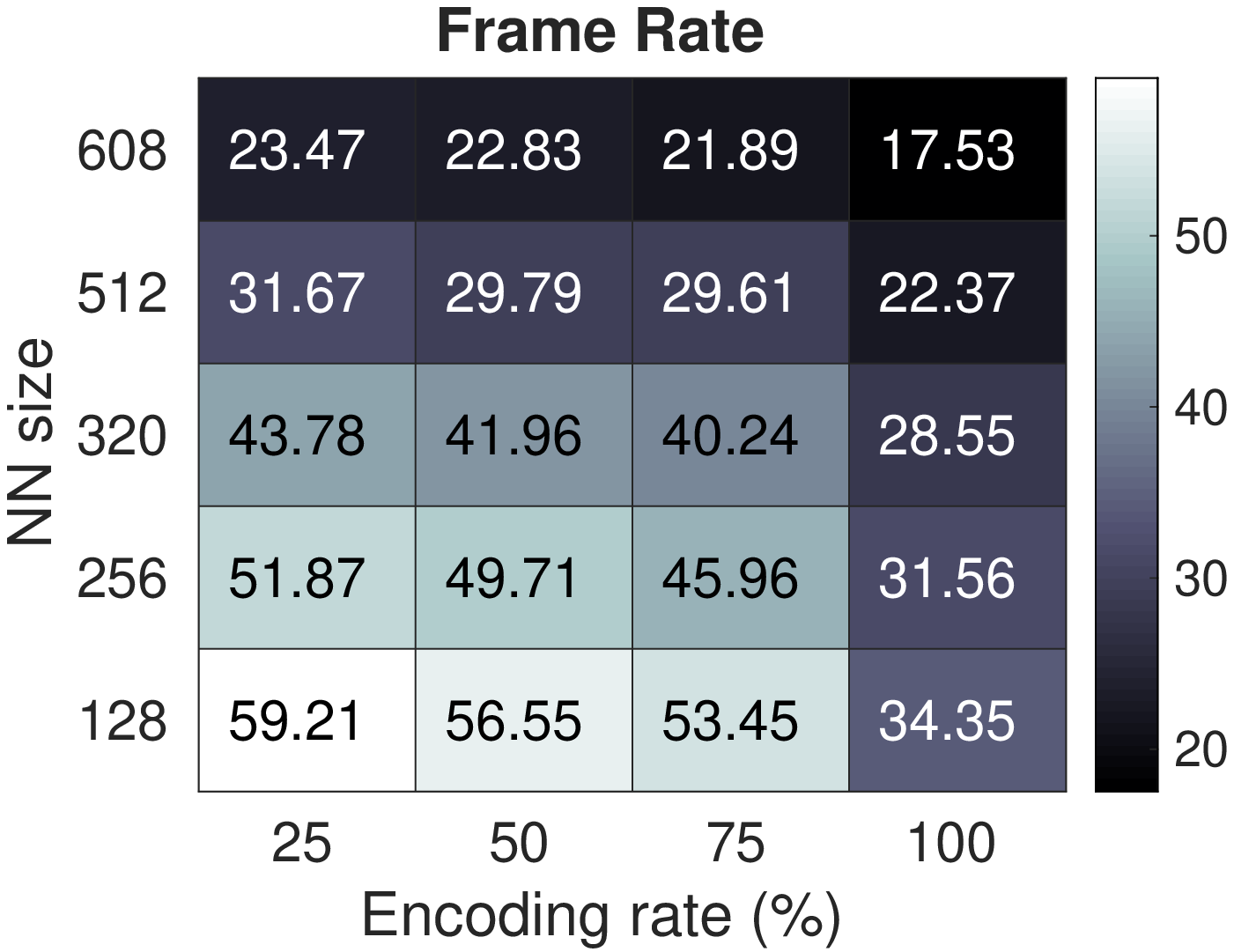}
	\caption{}
	\label{fig:frame_rate_udp}
	\end{subfigure}
	\caption{Performance Trade-offs (disabled power-save; UDP). (a-b) Precision and recall (IoU=0.5) vs $n$ and $q$. (c-d) Precision and recall for medium and large objects vs $n$ for uncompressed images (in (c)), and vs $q$ with fixed $n=512$ (in (d)). (e) Frame rate vs NN size $n$ and encoding rate $q$. All results are averaged over all images and all IoUs in $[0.5, 0.95]$.	} 
	\vspace{-2mm}
\end{figure*}


%

\section{Performance Trade-offs} \label{sec:trade-offs}

Using our measurements above we discuss here the interaction and trade-offs between the two performance metrics, i.e., the accuracy and E2E delay, under a range of different scenarios. We discover that \emph{in several cases there are sharp trade-off curves which create opportunities for improving the system operation, by carefully tuning parameters $q$ and $n$}.

Figures \ref{fig:precision}-\ref{fig:recall} plot the object recognition\footnote{We have used the Python library CoCoApi for calculating these metrics, \url{https://github.com/cocodataset/cocoapi/tree/master/PythonAPI/pycocotools}.} average precision and recall vs the encoding rate $q$ and the NN size $n$. We see that both metrics generally increase with $q$ and $n$, although there is a sharp improvement going from $n\!=\!128$ to $n\!=\!256$. Moreover, as $n$ drops the precision and recall performance deteriorate and cannot be improved even if we use high $q$ (e.g., see last row in each matrix). This finding differs from previous studies, e.g., \cite{dodge2016}, perhaps due to the COCO dataset which contains images with a large range of object sizes.

We further study the impact of the object sizes on performance, while we consider different detection thresholds (IoU values) \cite{COCO}. In Fig.~\ref{fig:accuracy_vs_N} we plot the precision and recall vs $n$ and $q$ for large and medium objects, averaged for a range of IoU values. We see that for large objects the accuracy increases rapidly with $n$ but plateaus when $n\!>\!300$. For medium objects on the other hand, the benefits of larger input size (and so higher image resolution) are greater and accuracy only plateaus when $n\!>\!500$. Fig.~\ref{fig:accuracy_vs_Q} shows that the dependence on $q$, albeit not that strong, follows indeed a continuous increase. We note that the precision and recall values in these plots are relatively low because we use very high IoU thresholds (up to 0.95). Also, we do not consider larger NNs since for $n\!=\!608$ we already have satisfactory precision but also large delays.



Finally, we study the frame rate, i.e., the reciprocal of E2E latency, for different NN sizes and image encoding rates. Fig.~\ref{fig:frame_rate_udp} presents the average frame rate for each scenario. Notice that for small NNs ($n\!<\!320$) the encoding affects significantly the frame rate, but this effect is weaker for $n\!>\!320$. For example, when $n\!=\!608$ the rate falls below 30fps even for very small values of $q$. In other words, we find that in the low NN size regime, the accuracy gains from choosing a high encoding rate are not significant, while the frame rate gains of a low encoding rate are substantial. Hence, a low encoding rate is probably more suitable for a small NN. The opposite is true in the high NN size regime, where we can achieve substantial accuracy gains without compromising significantly the frame rate. \emph{These findings underline the importance of selecting jointly the values of parameters $n$ and $q$}. Next section provides a systematic methodology towards that end.

\section{Data Models and Pareto Analysis} \label{sec:optimiz}



\subsection{Fitting the Measurements}

Our measurements indicate that the latency components and accuracy can be approximated using quadratic functions of the decision variables $n$ and $q$. Note that only the decoding delay $T_{dec}$ and precision $f$ (we omit recall for brevity) depend on both $n$ and $q$. On the other hand, the encoding and transmission delays, $T_{enc}$ and $T_{tx}$, depend only on $q$, and the deep learning delay $T_{dl}$ on $n$. We therefore define:
\begin{align}
	T_{enc}(q) &= \alpha_0 + \alpha_1 q + \alpha_2 q^2, \label{eq:T_enc} \\
	T_{dec}(n,q) &= \beta_0 + \beta_1 n + \beta_2 q + \beta_3 n q + \beta_4 n^2 + \beta_5 q^2, \label{eq:T_dec} \\
	T_{tx}(q) &= \gamma_0 + \gamma_1 q + \gamma_2 q^2, \label{eq:T_tx} \\
	T_{dl}(n) &= \delta_0 + \delta_1 n + \delta_2 n^2, \label{eq:T_dl} \\
	f(n,q) &= \epsilon_0 + \epsilon_1 n + \epsilon_2 q + \epsilon_3 n q + \epsilon_4 n^2 + \epsilon_5 q^2. \label{eq:f_acc}
\end{align}
The model parameters are obtained by fitting our measurements to \eqref{eq:T_enc}-\eqref{eq:f_acc}. Clearly, the exact values of these parameters can change if, for instance, we use a different access point or server. However, as our tests with the different handset devices have revealed, the changes are minimal and only quantitative.\footnote{The handsets affect only the values of parameters $\{\alpha_i\}_{i}$ and $\{\gamma_i\}_i$.}  


\subsection{Pareto Analysis}

We leverage the above models to explore the interaction of the decision variables:
\begin{equation}
n\in\mathcal{N}\!\triangleq\!\big\{[128,608]\mid\text{mod}(n,32)\!=\!0\big\},\,\,\,\,q \in \mathcal{Q}\!\triangleq\![10,100], \notag
\end{equation} 
i.e., study how they jointly affect the precision and the frame rate (E2E latency), while we also devise the Pareto fronts for these two performance criteria by following a detailed parameter-sensitivity analysis. We formulate two optimization problems; $\mathbb P_1$, where we maximize the precision subject to achieving a minimum frame rate; and $\mathbb P_2$ where we maximize the frame rate while not dropping the precision below a threshold value. Formally the 2 problems can be written:

\begin{mdframed}[innertopmargin=1mm,innerbottommargin=1mm,skipabove=1mm,skipbelow=1mm]
\begin{align}
\mathbb P_1:\quad	\underset{n\in \mathcal{N},q \in \mathcal{Q}}{\text{maximize}}\ &\,\,f(n,q) \label{eq:objective}\\
	\text{s.t.}\ &\,\, T_{total}(n,q) \leq T_{max} \label{eq:constraint}
\end{align} 
\end{mdframed}
\vspace{-2mm}
\begin{mdframed}[innertopmargin=1mm,innerbottommargin=1mm,skipabove=1mm,skipbelow=1mm]
\begin{align}
\mathbb P_2:\quad	\underset{n\in \mathcal{N},q \in \mathcal{Q}}{\text{minimize}}\ &\,\, T_{total}(n,q) \label{eq:objective_2}\\
	\text{s.t.}\ &\,\, f(n,q) \geq f_{min}. \label{eq:constraint_2}
\end{align} 
\end{mdframed}
where we have defined:
\begin{equation}
T_{total}(n,q) = T_{enc}(q)+T_{dec}(n,q)+T_{tx}(q)+T_{dl}(n),\notag
\end{equation}
and $T_{max}$ is the highest tolerable delay in order to achieve a frame rate of $1/T_{max}$ fps. Respectively, $f_{min}$ is the target precision requested by the user. In essence, constraint \eqref{eq:constraint} ensures that the total delay does not exceed $T_{max}$, and hence the frame rate $1/T_{total}$ will be greater or equal to the threshold $1/T_{max}$. Similarly in $\mathbb P_2$ we maximize the frame rate by minimizing $T_{total}$. Using both problem formulations we will be able to highlight the trade-offs between delay and precision. 

\begin{figure}[t] 
	\centering
	\begin{subfigure}{.48\linewidth}
		\includegraphics[scale=0.31]{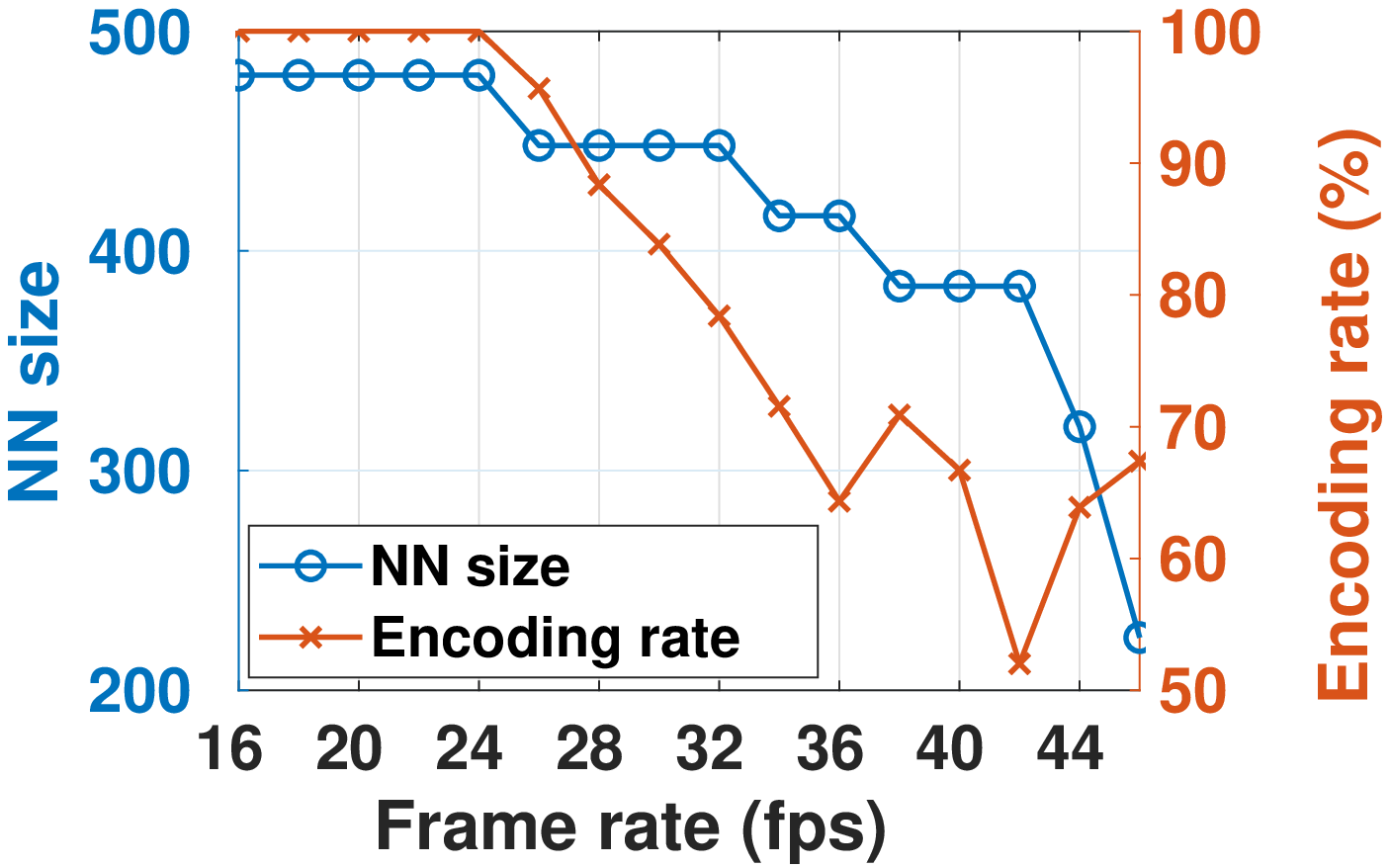}
		\caption{}\label{fig:opt1_paras}
	\end{subfigure}
	\begin{subfigure}{.48\linewidth}
		\includegraphics[scale=0.31]{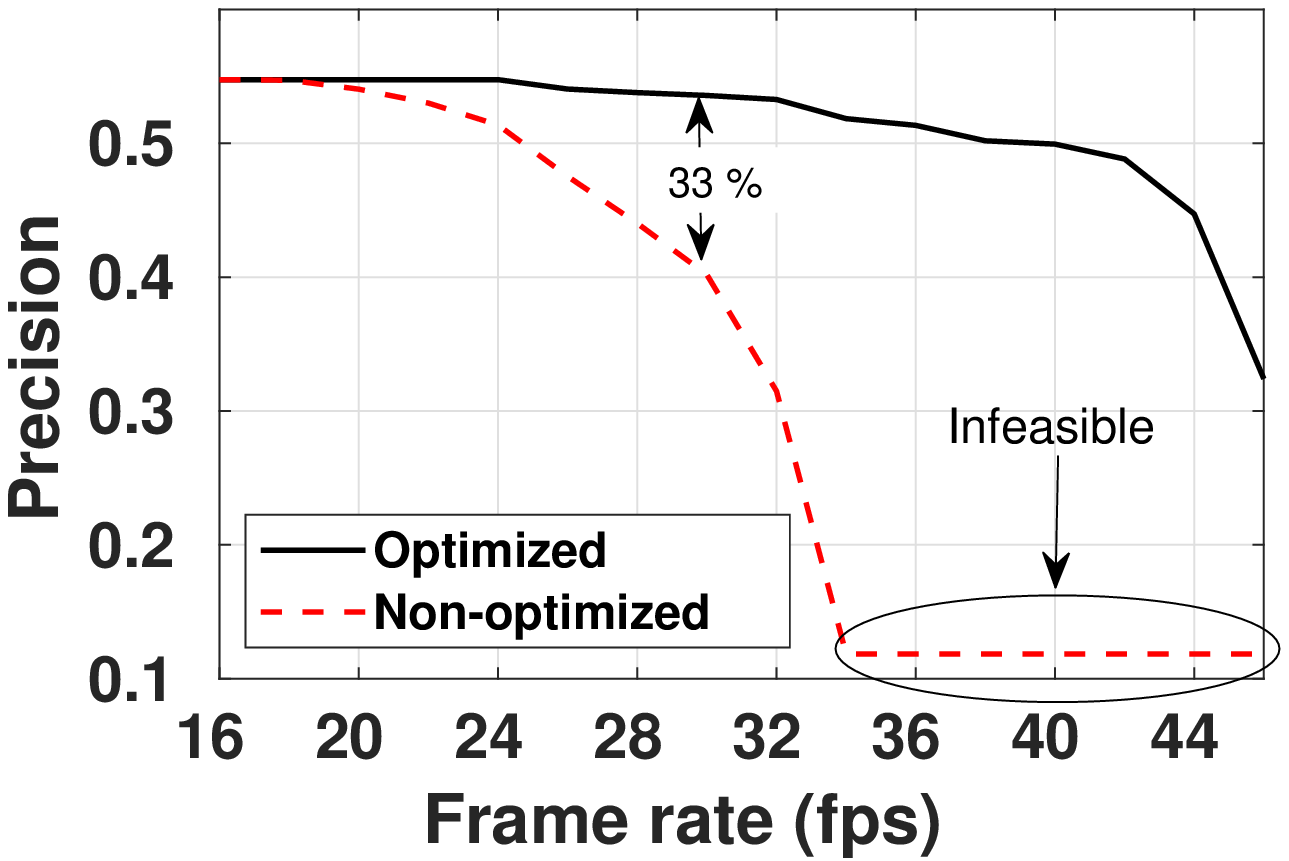}
		\caption{}\label{fig:opt1_p}
	\end{subfigure}
	\caption{(a) Optimal NN size and encoding rate for the desired frame rate. (b) Corresponding maximal precision values. }
	\vspace{-2mm}
\end{figure}

Fig.~\ref{fig:opt1_paras} plots the values of $n$ and $q$ that maximize the precision while keeping the frame rate at or above the value indicated on the x-axis (recall that $n$ is a multiple of 32). The achieved precision for each frame rate is displayed with a solid line in Fig.~\ref{fig:opt1_p}. Observe how the increasing frame rate dictates the drop of NN size and encoding rate, which in turn result in decreasing precision performance. Moreover, we observe that the NN size continuously drops or stays level with the frame rate, while the encoding rate can increase in some cases. That occurs when the NN size has been reduced and hence the increase of the encoding rate can sustain a higher precision. Notice that for the largest range of frame rates, the NN size can be kept quite high (around and above 400), even when exceeding 30 fps. This yields a satisfactory precision of $0.5$ at 40 fps\footnote{Recall that we obtain low precision values because on purpose we used very high IoU values; for more typical thresholds the precision is much higher.}. However, after the 40 fps threshold, the NN size has to be very small to facilitate fast object recognition and the precision performance drops dramatically.

\begin{figure}[t]
	\centering
	\begin{subfigure}{.48\linewidth}
		\includegraphics[scale=0.3]{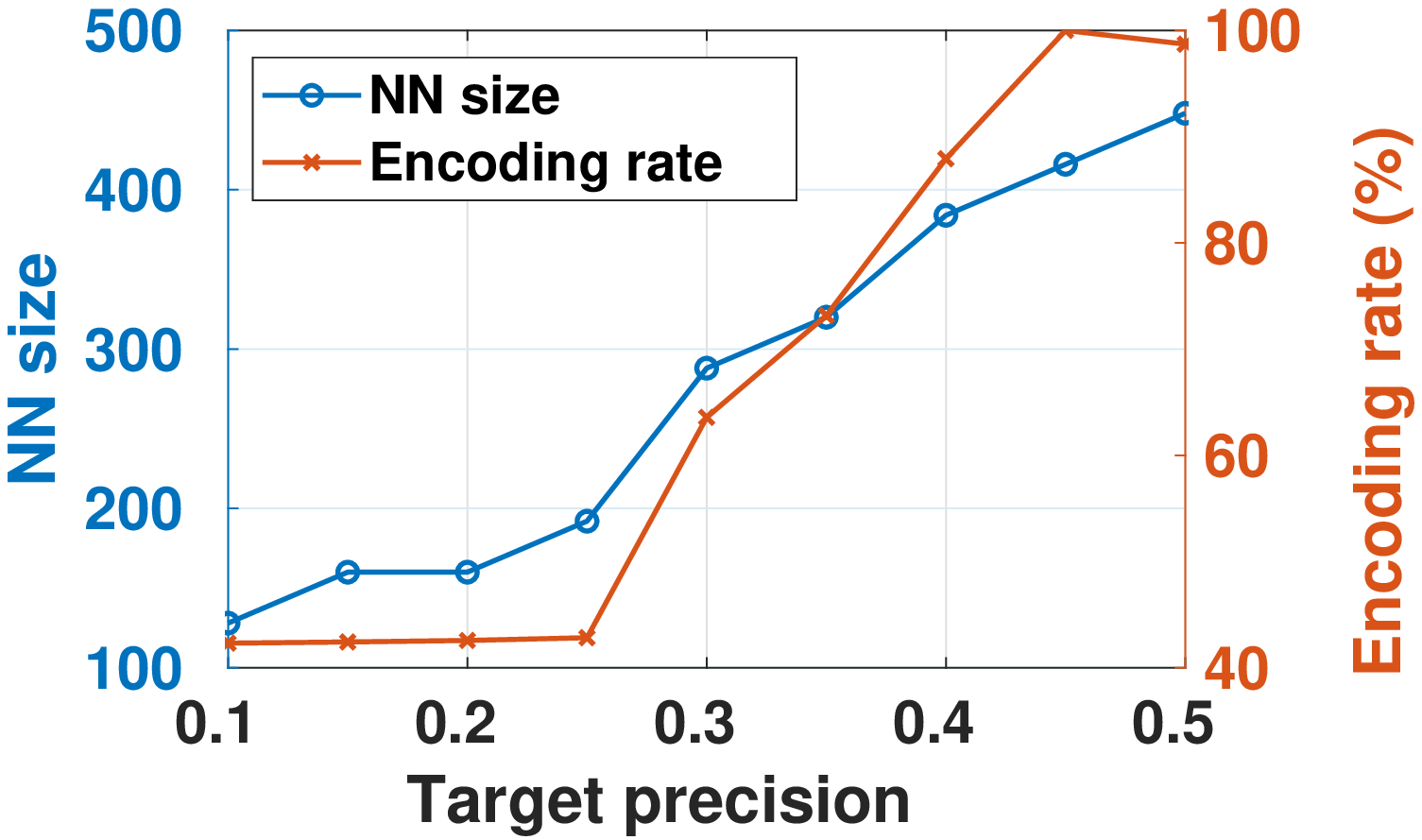}
		\caption{}\label{fig:opt_vars}
	\end{subfigure}
	~
	\begin{subfigure}{.48\linewidth}
		\includegraphics[scale=0.3]{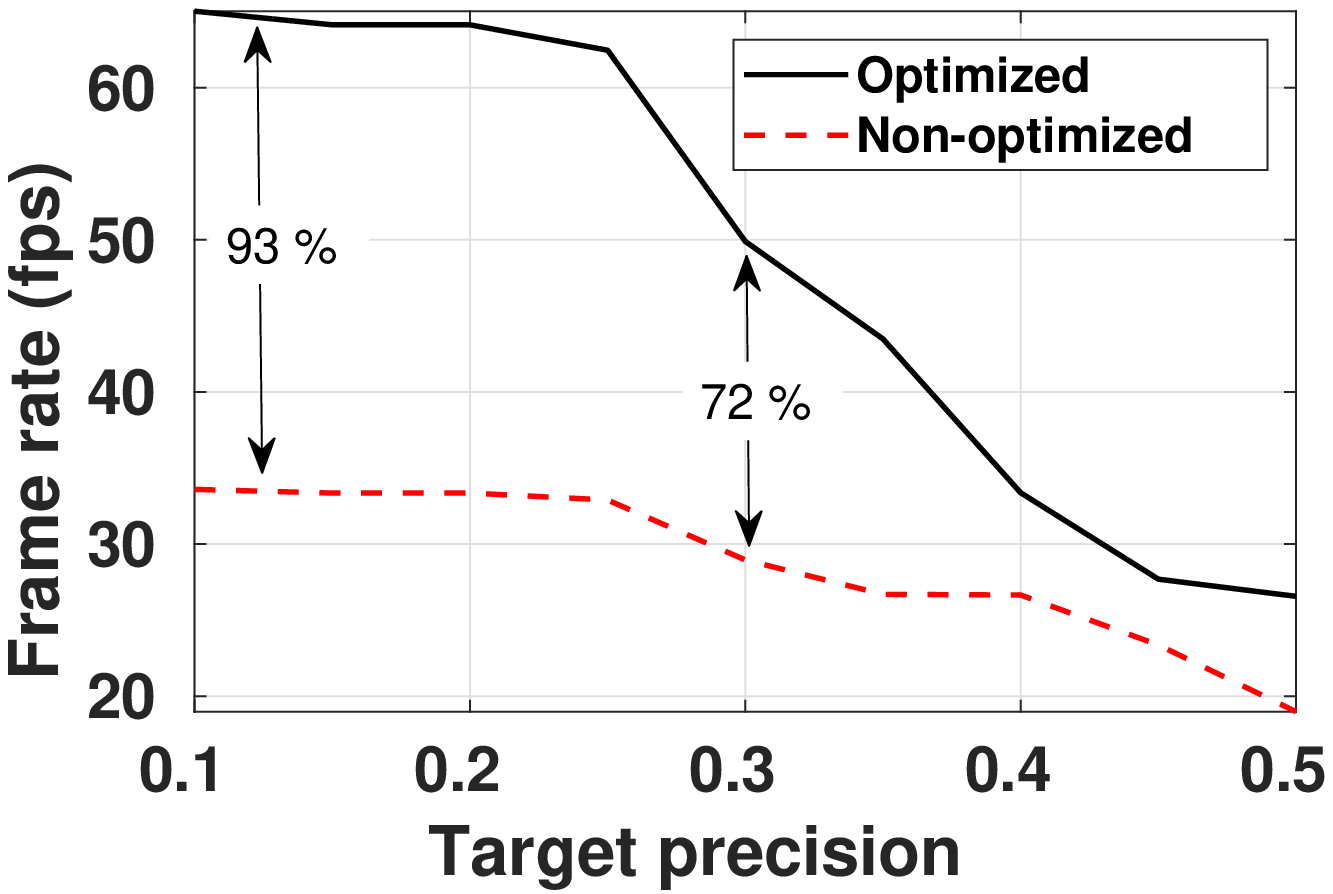}
		\caption{}\label{fig:opt_fr}
	\end{subfigure}
	\caption{(a) Optimal NN size and encoding rate for the target accuracy. (b) Corresponding maximal frame rate values.}
	\vspace{-2mm}
\end{figure}

To highlight the impact of our optimized networking configuration, we compare the performance with the respective results of a non-optimized (\emph{vanilla}) system, dashed line in Fig.~\ref{fig:opt1_p}. Namely, these results were obtained by fitting the non-optimized (TCP, and enabled powersave) wireless transmission delay measurements to \eqref{eq:T_tx} and solving $\mathbb P_1$. Clearly, the increased transmission delays hamper the ability of the system to achieve high precision for acceptable frame rates (precision drops by 33\% at 30 fps). Moreover, $\mathbb P_1$ becomes infeasible for a target frame rate above 34 fps, indicating the greater range in which the system can operate after configuring the network. The respective results for $\mathbb P_2$ are displayed in Fig.~\ref{fig:opt_vars},~\ref{fig:opt_fr}. The optimal frame rate can be kept very close to 30 fps, even for very high target precision. Also, we observe a huge gap between the optimized and non-optimized solution in this case, with the former achieving up to 93\% higher frame rate than the latter when target precision is very low.


\section{Related Work} \label{sec:related}

\textbf{Deep Learning With Compressed Images.}
The impact of image compression on recognition accuracy has started to receive attention in the deep learning literature, see seminal paper~\cite{dodge2016} and follow-up works, but this aspect of performance remains relatively poorly understood.  Most attention has focused on developing new compression approaches tailored to deep learning e.g. see \cite{DeepN,edge_assisted}. The authors in~\cite{Zhang18} explore the effect of image compression rate to the object detection accuracy. To the best of our knowledge however, the system-level trade-offs between E2E latency \emph{and} deep learning accuracy introduced by the use of image compression have not been previously explored. 

\textbf{Edge-Assistance.}
JAGUAR~\cite{JAGUAR} and Glimpse~\cite{Glimpse} are edge-assisted, real-time object recognition systems. They both use object tracking to reduce the number of recognitions, but do not use state of the art deep learning techniques for object recognition.
\cite{DL_offloading} proposes a solution for deciding the execution location of augmented reality tasks, either on the mobile, or an edge server. The idea of distributing the neural network layers among different tiers of the network architecture is demonstrated in~\cite{DDNN_1,DDNN_2}. The devices, based on their computation resources execute smaller or larger parts of the NN towards increasing the accuracy of inferences with tolerable execution and network delays. In~\cite{AAY18} the authors propose a framework for distributing deep learning sub-processes to edge, cloudlet and cloud nodes towards increasing the job execution rate of the system. 
\cite{edge_assisted} presents an augmented reality object detection system that leverages an edge server, as well as object tracking and image encoding to improve latency. 
The above works indicate the necessity of edge architectures, towards improving the E2E latency of delay sensitive services.

\textbf{Accuracy/Latency Trade-off.}
JALAD~\cite{JALAD} proposes the decoupling of a Deep Neural Network (DNN) between edge and cloud towards minimizing latency with execution accuracy guarantees. Overlay~\cite{Overlay} presents an augmented reality system for mobile devices, assisted by a GPU-enabled server that is designed towards minimizing the tracking error. MobiQoR~\cite{MobiQoR} studies the trade-off between delay and Quality of Result for edge applications that involve machine learning and analytics like face recognition. The authors show that sacrificing computation result quality can decrease delay as well as energy consumption. 
LAVEA~\cite{LAVEA} proposes a system for computation offloading of data analytics to nearby edge nodes. The formulated optimization problem aims in making offloading and bandwidth allocation decisions towards minimizing latency. 
%
DeepDecision~\cite{DeepDecision} is a video analytics system that balances accuracy and latency, by properly adjusting the camera sample rate, video encoding rate, and deep learning model. However, both transmission and processing delays are much higher than the ones obtained by our system.
DeepMon~\cite{DeepMon} distributes the execution of a large DNN across multiple mobile GPUs to reduce latency. It focuses on DNN optimizations, instead of the network-centric analysis presented in our work.
All the above works, highlight the inherent trade-off between latency and accuracy in edge architectures. Our work however goes beyond that, by proposing important delay reducing modifications that easily enable real time performance for object recognition.



\section{Conclusions} \label{sec:conclusion}
We develop an edge-assisted object recognition system and show that careful network transmit and powersave strategies can significantly reduce the wireless transmission delay. We find that the level of image compression, as well as the dimension of the deep learning network used, are key design parameters, affecting both end-to-end latency and object recognition accuracy. We demonstrate how our measurements can be used to choose these design parameters to optimally trade-off between execution delay and accuracy. 

\section*{Acknowledgments}
The authors would like to thank Domenico Guistianino for helpful input and discussions during development of the system. This publication has emanated from research supported in part by SFI research grants 17/CDA/4760, 16/IA/4610 and is co-funded under the European Regional Development Fund under Grant Number 13/RC/2077.

\bibliographystyle{IEEEtran}
\bibliography{citations}

\begin{thebibliography}{10}
\providecommand{\url}[1]{#1}
\csname url@samestyle\endcsname
\providecommand{\newblock}{\relax}
\providecommand{\bibinfo}[2]{#2}
\providecommand{\BIBentrySTDinterwordspacing}{\spaceskip=0pt\relax}
\providecommand{\BIBentryALTinterwordstretchfactor}{4}
\providecommand{\BIBentryALTinterwordspacing}{\spaceskip=\fontdimen2\font plus
\BIBentryALTinterwordstretchfactor\fontdimen3\font minus
  \fontdimen4\font\relax}
\providecommand{\BIBforeignlanguage}[2]{{%
\expandafter\ifx\csname l@#1\endcsname\relax
\typeout{** WARNING: IEEEtran.bst: No hyphenation pattern has been}%
\typeout{** loaded for the language `#1'. Using the pattern for}%
\typeout{** the default language instead.}%
\else
\language=\csname l@#1\endcsname
\fi
#2}}
\providecommand{\BIBdecl}{\relax}
\BIBdecl

\bibitem{killer_app}
G.~{Ananthanarayanan} \emph{et~al.}, ``Real-time video analytics: The killer
  app for edge computing,'' \emph{Computer}, vol.~50, no.~10, pp. 58--67, 2017.

\bibitem{Zhang18}
J.~{Ren}, Y.~{Guo}, D.~{Zhang}, Q.~{Liu}, and Y.~{Zhang}, ``Distributed and
  efficient object detection in edge computing: Challenges and solutions,''
  \emph{IEEE Network}, vol.~32, no.~6, pp. 137--143, 2018.

\bibitem{edge_assisted}
L.~Liu, H.~Li, and M.~Gruteser, ``Edge assisted real-time object detection for
  mobile augmented reality,'' in \emph{Proc. of ACM MobiCom}, 2019.

\bibitem{Glimpse}
T.~Y.-H. Chen \emph{et~al.}, ``Glimpse: Continuous, real-time object
  recognition on mobile devices,'' in \emph{Proc. of ACM SenSys}, 2015.

\bibitem{dodge2016}
S.~{Dodge} and L.~{Karam}, ``Understanding how image quality affects deep
  neural networks,'' in \emph{QoMEX}, 2016.

\bibitem{DeepMon}
L.~N. Huynh, Y.~Lee, and R.~K. Balan, ``Deepmon: Mobile gpu-based deep learning
  framework for continuous vision applications,'' in \emph{Proc. of ACM
  MobiSys}, 2017.

\bibitem{DeepDecision}
X.~{Ran} \emph{et~al.}, ``Deepdecision: A mobile deep learning framework for
  edge video analytics,'' in \emph{Proc. of IEEE INFOCOM}, 2018.

\bibitem{AAY18}
M.~{Ali} \emph{et~al.}, ``Edge enhanced deep learning system for large-scale
  video stream analytics,'' in \emph{Proc. of IEEE ICFEC}, 2018.

\bibitem{Overlay}
P.~Jain, J.~Manweiler, and R.~Roy~Choudhury, ``Overlay: Practical mobile
  augmented reality,'' in \emph{Proc. of ACM MobiSys}, 2015.

\bibitem{yolov3}
J.~Redmon and A.~Farhadi, ``Yolov3: An incremental improvement,'' \emph{arXiv},
  2018.

\bibitem{DL_offloading}
X.~Ran, H.~Chen, Z.~Liu, and J.~Chen, ``Delivering deep learning to mobile
  devices via offloading,'' in \emph{Workshop on Virtual Reality and Augmented
  Reality Network}, 2017.

\bibitem{LAVEA}
S.~Yi \emph{et~al.}, ``Lavea: Latency-aware video analytics on edge computing
  platform,'' in \emph{Proc ACM/IEEE Symposium on Edge Computing}, 2017.

\bibitem{COCO}
T.~Lin \emph{et~al.}, ``Microsoft {COCO:} common objects in context,''
  \emph{arXiv}, vol. abs/1405.0312, 2014.

\bibitem{DeepN}
Z.~Liu \emph{et~al.}, ``Deepn-jpeg: {A} deep neural network favorable
  jpeg-based image compression framework,'' \emph{CoRR}, vol. abs/1803.05788,
  2018.

\bibitem{JAGUAR}
W.~Zhang, B.~Han, and P.~Hui, ``Jaguar: Low latency mobile augmented reality
  with flexible tracking,'' in \emph{Proc. of ACM Conference on Multimedia},
  2018.

\bibitem{DDNN_1}
C.~{Lo}, Y.~{Su}, C.~{Lee}, and S.~{Chang}, ``A dynamic deep neural network
  design for efficient workload allocation in edge computing,'' in \emph{Proc.
  of IEEE ICCD}, 2017.

\bibitem{DDNN_2}
S.~{Teerapittayanon}, B.~{McDanel}, and H.~T. {Kung}, ``Distributed deep neural
  networks over the cloud, the edge and end devices,'' in \emph{Proc. of IEEE
  ICDCS}, 2017.

\bibitem{JALAD}
H.~Li \emph{et~al.}, ``{JALAD:} joint accuracy- and latency-aware deep
  structure decoupling for edge-cloud execution,'' \emph{CoRR}, vol.
  abs/1812.10027, 2018.

\bibitem{MobiQoR}
Y.~{Li}, Y.~{Chen}, T.~{Lan}, and G.~{Venkataramani}, ``Mobiqor: Pushing the
  envelope of mobile edge computing via quality-of-result optimization,'' in
  \emph{Proc. of IEEE ICDCS}, 2017.

\end{thebibliography}

\end{document}